\documentclass[11pt,twoside]{article}

\usepackage{asp2006}

\begin{document}

\def\teff{$T\rm_{eff }$}
\def\kms{$\mathrm {km s}^{-1}$}
\def\cntrt{counts cm$^{-2}$s$^{-1}$keV$^{-1}$} 
\def\sgra{Sgr~A$^*$}
\def\Msol{M$_{\odot}$}
\def\lum{10$^{35}$ erg s$^{-1}$}

\title{An overview of the high-energy emission from the Galactic Center}   
\author{Andrea Goldwurm}   
\affil{Service d'Astrophysique/IRFU/CEA-Saclay, F-91191 Gif-sur-Yvette $\&$
AstroParticule et Cosmologie, Universit{\' e} Paris 7, F-75205 Paris, France}

\begin{abstract} 
The Galactic Center is a prominent source in X-rays and gamma-rays
and the study of its high-energy emission is crucial in understanding the
physical phenomena taking place in its dense and extreme environment,
phenomena that are possibly common to other galactic nuclei.
However this emission is also very complex and consists of both thermal 
and non thermal radiation produced by compact and extended sources, 
surrounded by more diffuse components. In spite of the 
fundamental advances obtained in the last ten years with Chandra, XMM-Newton, 
INTEGRAL, HESS and Suzaku several questions remain open to investigations.
I will review here the main results and the open issues on
the high-energy diagnostic of the galactic nuclear activity.
\end{abstract}
%

\section{Introduction}
The Galactic Center (GC), the sky region of $\sim$ 4$^\circ$ $\times$ 2$^\circ$ in
galactic longitude and latitude respectively, 
surrounding the center of our Galaxy, and
corresponding to about 600~pc $\times$ 300~pc size at the estimated distance of 8~kpc,
has been observed with high-energy instruments since the very
beginning of the X/gamma-ray astronomy.
One of the major motivations of these surveys has been to measure  
the energetic radiation from the GC super-massive black hole (SMBH)
that is now firmely detected (with an estimated mass of 4~10$^{6}$~M$_{\odot}$)
and associated to the compact radio source \sgra.
However several other interesting and in fact much brighter sources populate the region: 
variable X-ray binaries (XRB), 
supernova remnants (SNR) that interact with dense molecular clouds (MC), stellar clusters 
of young stars creating HII regions and emitting powerful stellar winds, 
non-thermal filaments, pulsar wind nebulae and a hot gas mixed with energetic particles.
Totally obscured in the optical and ultraviolet wavelengths by the galactic plane dust, 
the GC is mainly observed from radio to infrared (IR) frequencies and then 
at high energies ($>$~1~keV). 
A radio picture of the region, that shows its complexity, is reported in Fig. 1
and general reviews are given by Mezger et al. 1996; Morris $\&$ Serabyn 1996; 
Melia $\&$ Falcke 2001; Melia 2007.

The high-energy data are particularly interesting to trace the most violent 
phenomena generated by strong gravitational or magnetic fields, and that give rise 
to particle acceleration and non-thermal radiation.
In the recent years, with the launch of Chandra, XMM-Newton, INTEGRAL and Suzaku
X-ray/gamma-ray observatories and with the operation of ground-based very-high-energy (VHE) 
gamma-ray observatories like HESS, this quest has led to some fundamental discoveries.
I will summarise here these results, focussing on items other than \sgra\ emission, and 
updating my previous review works on similar subject (Goldwurm 2006, 2007, 2009) by
reporting the most recent findings and discussing the still open questions.

\section{Early X-ray and gamma-ray observations}

The detection of high-energy emission from the GC dates back to the 
very beginning of the X-ray astronomy, in the 1960s, with devices mounted on sounding rockets.
The first claim of detection of a galactic center X-ray source in Sagittarius
appeared in a paper by Bowyer et al. (1965) of the Naval Research laboratory (NRL), 
reporting results from an Aerobee rocket flight in 1964. 
Several other measurements followed by the AS$\&$E, MIT, Lockeed and 
again the NRL X-ray astronomy groups \cite{cla65,fis66,gur67,bra68}. 
From these first investigations appeared that most of
the newly discovered X-ray sources were clustering around the GC 
but the nucleus itself did not seem bright.
In the 70s the region was monitored by the first X-ray satellites, Uhuru, Ariel~5 and HEAO~1
and other rocket experiments.
The GC X-ray source (4U~1743-29) detected by Uhuru in 1970 appeared somewhat extended 
and either due to diffuse emission or to a composition of several sources \cite{kel71} while
Ariel 5 and other experiments revealed the presence of bright variable or
transient objects (e.g. A~1742-294 and A~1742-289) and of several burst sources 
that were positioned with large error boxes.
At the eve of the launch of the Einstein Observatory it was already clear \cite{pro78,cru78}
that, in spite of the large activity of the GC region, the X-ray luminosity of the nucleus 
was much lower ($<$ few 10$^{36}$ erg s$^{-1}$) than in Active Galactic Nuclei (AGN).

The first GC X-ray images with arcminute resolution were however obtained 
only when, at the end of the 70s with Einstein (HEAO~2), it was possible to implement
focussing mirrors for soft X-ray telescopes. Watson et al. (1981) showed then 
that the central 20$'$ of the Galaxy at $<$~4~keV were dominated by diffuse emission with 
several weak point-like sources, one of which associated to Sgr~A~West and
therefore including \sgra.
This object was then resolved in 3 weaker sources with Rosat \cite{pretru94}
more then 10 years later leading to a measurement of soft X-ray luminosity of only
10$^{34}$~erg~s$^{-1}$ for the source coincident (within 10$''$) with the nucleus. 
On the other hand the GC SMBH could still shine in hard X-rays or even at 511~keV, 
the positron-electron annihilation line, since important, variable, fluxes were observed 
from the general direction of the GC since the late 60s at these energies. 
After all, BH binaries, like Cyg X-1, often emit the bulk of their accretion 
luminosity at $>$~100 keV. 
Hard X-ray observations of the GC started nearly as early as X-ray observations 
using NaI detectors on stratospheric balloons and led to significant detections of both
hard (30-150 keV) continuum and 500~keV line from the GC (Haymes et al. 1969, 1975) 
with localisation uncertainties of few degrees. 
Further measurements confirmed in particular the presence of a GC 511~keV narrow 
and variable emission
line associated to an orthopositronium continuum (e.g. Lingenfelter $\&$ Ramaty 1989).

Untill the 90s however it was not possible, due to the uncertainties in these measurements,
to disantangle the different components and these detections
were still quoted as evidences for the presence of a massive BH at the galactic center 
(see e.g Genzel $\&$ Townes 1987).
Coded mask imaging technics operated a similar revolution in the hard X-ray and soft gamma-ray band 
as the one operated by the grazing incidence X-ray mirrors in the soft X-ray regime,
by increasing angular resolution of hard-X/soft-gamma-ray telescopes to better than 30 arcmin.
Observations in hard X-rays (3-30 keV) with coded mask telescopes XRT/SpaceLab2 \cite{ski87} 
and ART-P/GRANAT \cite{pav94} and then in soft gamma-rays (30-1000 keV) 
with SIGMA/GRANAT \cite{gol94,gol01} showed then that the GC hard emission was rather 
due to the powerful hard XRBs of the region than to the nucleus.
In particular 1E~1740.7--2942, one of the unpeculiar sources discovered with Einstein at about 40$'$ 
from \sgra, was seen to fully dominate the images in those energy bands and was later 
recognized to be a very special object, the first BH microquasar. 
In addition, the SIGMA telescope set upper limits on the presence 
of a point-like source of 511 keV line \cite{mal95} 
while in those same years OSSE/CGRO showed that the bulk of the 511~keV line emission 
was not variable but rather constant, diffuse and extended over the whole 
galactic bulge \cite{pur97} and therefore not directly related to \sgra.
At higher energies ($>$~100 MeV), where coded mask are not efficient, 
the EGRET/CGRO telescope detected a GC 
gamma-ray source sticking out from the general gamma-ray emission 
produced by cosmic-ray interaction with the dense gas of the region \cite{may98}.
This source was positioned slightly away from the nucleus ($\sim$~0.2$^{\circ}$)  
but it could still be linked to \sgra\ given the large localisation error \cite{har99}.

Meanwhile a number of important results were also obtained on the GC X-ray diffuse emission.
The japonese satellite GINGA discovered a prominent 6.7~keV iron line 
diffuse emission from the region \cite{koy89}. 
The associated continuum component resembled to the diffuse galactic ridge emission (GRXE),
distributed along the galactic plane and characterised by a thin hot thermal plasma spectrum, 
discovered in the early 80s with HEAO~1 and EXOSAT missions \cite{war85}.
In the early 90s the ART-P telescope discovered a harder ($>$ 10 keV) component
associated to the molecular clouds of the region and 
this prompted the speculation that this emission could be due to reflection of 
X-rays emitted in the past by some very bright source, possibly \sgra\ itself \cite{sun93}.
The authors also pointed out that in this case a 6.4 keV line associated 
to the clouds shall also be detected. 
When the ASCA satellite in 1994 separated the diffuse contribution of the 6.7~keV line 
of ionized iron from the 6.4~keV one of neutral Fe atoms and showed that 
their distributions were different, with the 6.4~keV line one correlated to the molecular 
material, the hypothesis of a past flare from \sgra\ illuminating the clouds
was then explicitly formulated \cite{koy96}.

At the turn of the century, a new era of high-energy astronomy has been opened 
with the launch of the Chandra, XMM-Newton (1999) and Suzaku (2005) X-ray observatories; 
with the launch of the gamma-ray missions INTEGRAL (2002) and FERMI (2008) and the
start of operations of VHE gamma-ray ground telescopes like HESS (2003). 
\begin{figure}[!ht]
\plotfiddle{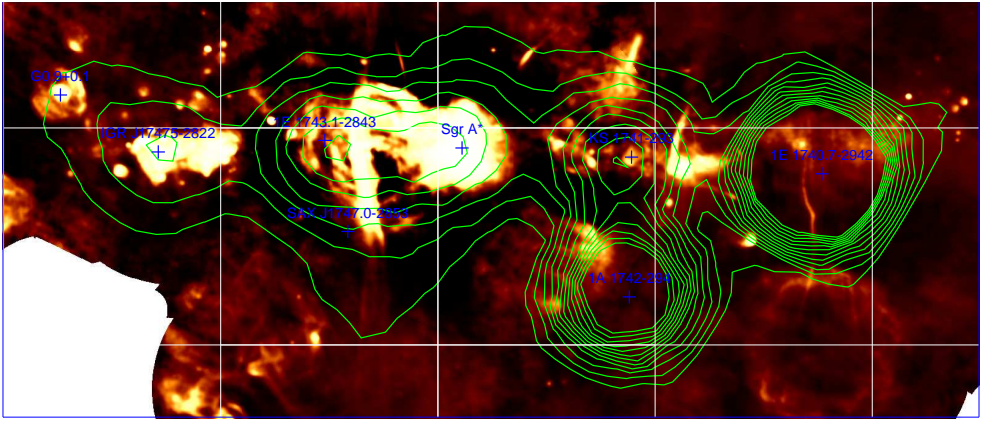}{6cm}{-90}{48}{48}{-200pt}{220pt}
\caption{
\footnotesize
Radio image of the galactic center at 90~cm from VLA data, 
on a grid of galactic coordinates spaced by 0.5$^{\circ}$ and compared to 
hard X-ray contours in the 20-40 keV band from the 2003-2004 INTEGRAL 
observations \cite{bel06}. 
}
\label{Radioimage}
\end{figure}

\begin{figure}[!ht]
\plotfiddle{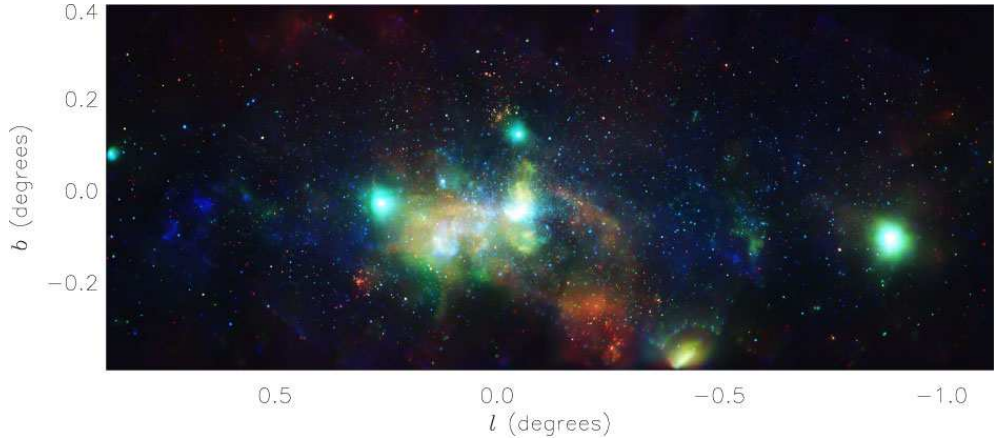}{6cm}{0}{67}{67}{-200pt}{-20pt}
\caption{
\footnotesize
X-ray image of the galactic center obtained with the deep survey of the
Chandra observatory in the 1--8~keV range (for the color figure: red is for 1--3~keV,
green for 3--5~keV and blue for 5--8~keV) \cite{mun09}.
}
\label{Xrayimage}
\end{figure}

\section{The Galactic Center in the X-ray band}

In the 1-10 keV X-ray band the GC is dominated by few bright, sometimes transient, 
X-ray binaries (e.g. 1E~1740.7--2942 or 1E~1743.1--2843).
The region also contains a large population of weak point-like persistent 
and transient sources, several components of diffuse emission, 
several peculiar sources of thermal and non-thermal radiation and
finally a weak central emission associate to \sgra\ (Fig.~2).

\subsection{The inner 3 pc: Sgr A* and its close environment}

Thanks to the exceptional resolution of the Chandra Observatory ($\sim$ 0.5$''$)
it has been possile to map in detail the central arcminute (2.4~pc)
of the galaxy in X-rays (Fig.~3 left). As for the whole GC, the X-ray morphology is quite different 
from the radio one, which is dominated here by the characteristic minispiral of Sgr~A West,
and from the IR one, dominated by the hot stars and by the dust associated
to the minispiral or the circumnuclear disk.
Several weak discrete X-ray sources were detected, surrounded by a diffuse emission \cite{bag03}. 

In the very center Chandra 
resolved the central Rosat source in several components and detected 
at the position of \sgra\ a very weak source with a steady 2-10~keV luminosity  
of only 2~10$^{33}$~erg~s$^{-1}$ (compared to the Eddington luminosity 
of $\sim$ 5~10$^{44}$~erg~s$^{-1}$ for the 4~10$^{6}$~M$_{\odot}$ SMBH), 
a steep spectrum and somehow extended over about 1$''$ \cite{bag03}. 
The quiescent \sgra\ emission is so low that it even challenges 
the radiative inefficient accretion flow models (RIAF) (Narayan et al. 1995, 
Yuan 2010) 
developped to explain the low luminosity of the galactic nucleus and 
in general the behavior of black holes in low accretion rate regime. 
The steep spectrum seems also little compatible with the hard thermal bremsstrahlung 
of such models, although the apparent extension of the source appears consistent with
an emission that starts at the Bondi radius ($\sim$ 10$^5$ R$_S$, where R$_S$
is the SMBH Schwarzschild radius $\sim$~1.2~10$^{12}$~cm or 1$''$ at the GC) 
as predicted by such theories.

The surrounding diffuse emission (within 15$''$ radius) has similar spectrum and likewise 
shows an iron line centered at 6.55 keV, which indicates fluorescence from plasma 
in non equilibrium ionization (NEI) \cite{xu06}.
This hot NEI plasma just around \sgra\ has been explained as due to the interaction
of the powerful stellar winds generated by the massive stars of the 
nearby young star clusters \cite{roc04,qua04}, stellar winds that shall also provide 
the bulk of the material accreting in the supermassive black hole.
A cometary source (G359.95-0.04) at about 10$''$ from \sgra\  
was also detected and interpreted as a pulsar wind nebula confined by ram pressure. 
In spite of its very weak X-ray emission it has been proposed \cite{wan06}
as candidate for the counterpart of the bright central TeV emission (sect. 4). 
The closeby IRS13 star cluster bright in infrared and supposed to host an 
intermediate-mass black hole, was also identified as another weak discrete object
of the central parsec.

\begin{figure}[!ht]
\plotfiddle{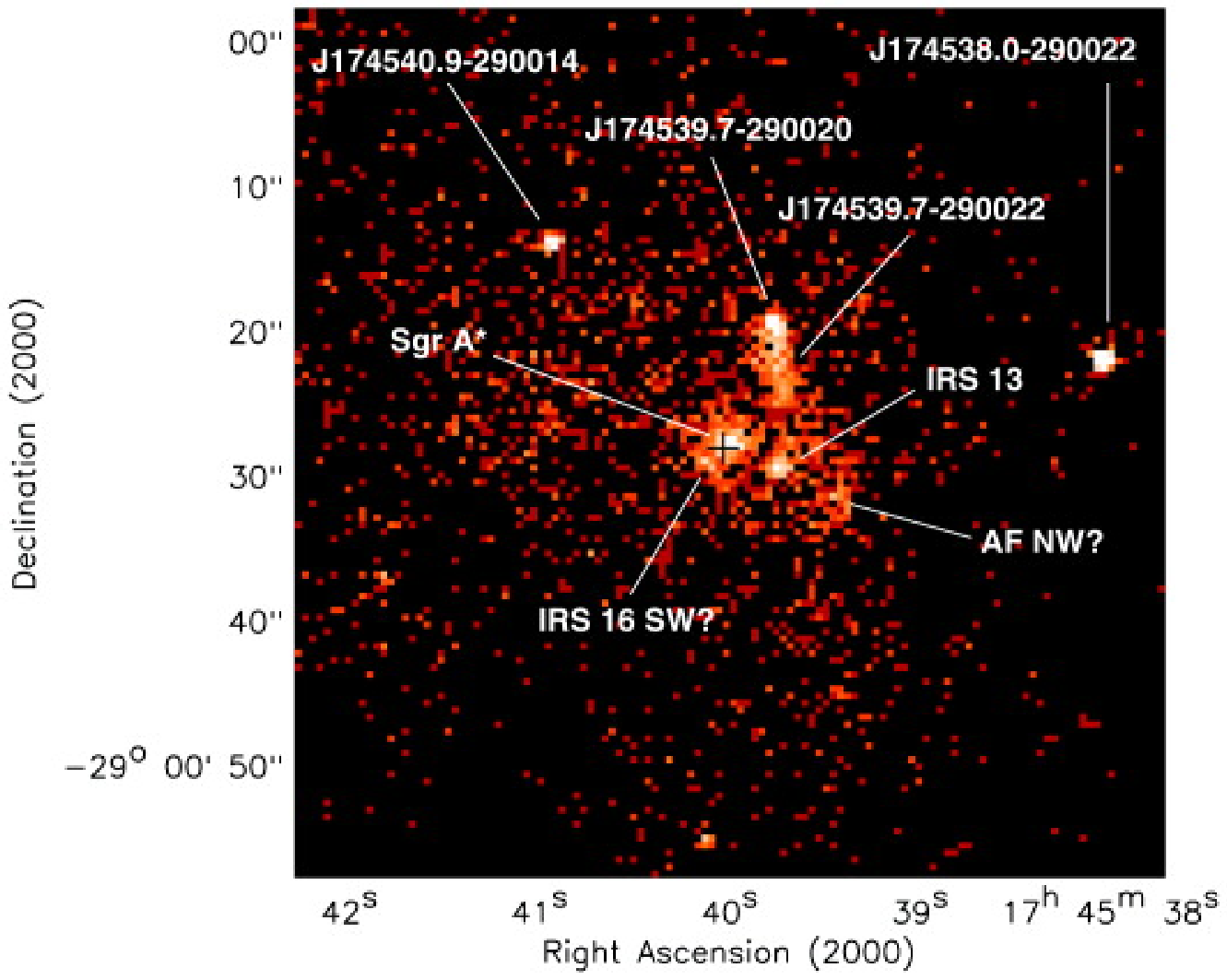}{2.5cm}{0}{49}{49}{-195pt}{-85pt}
\plotfiddle{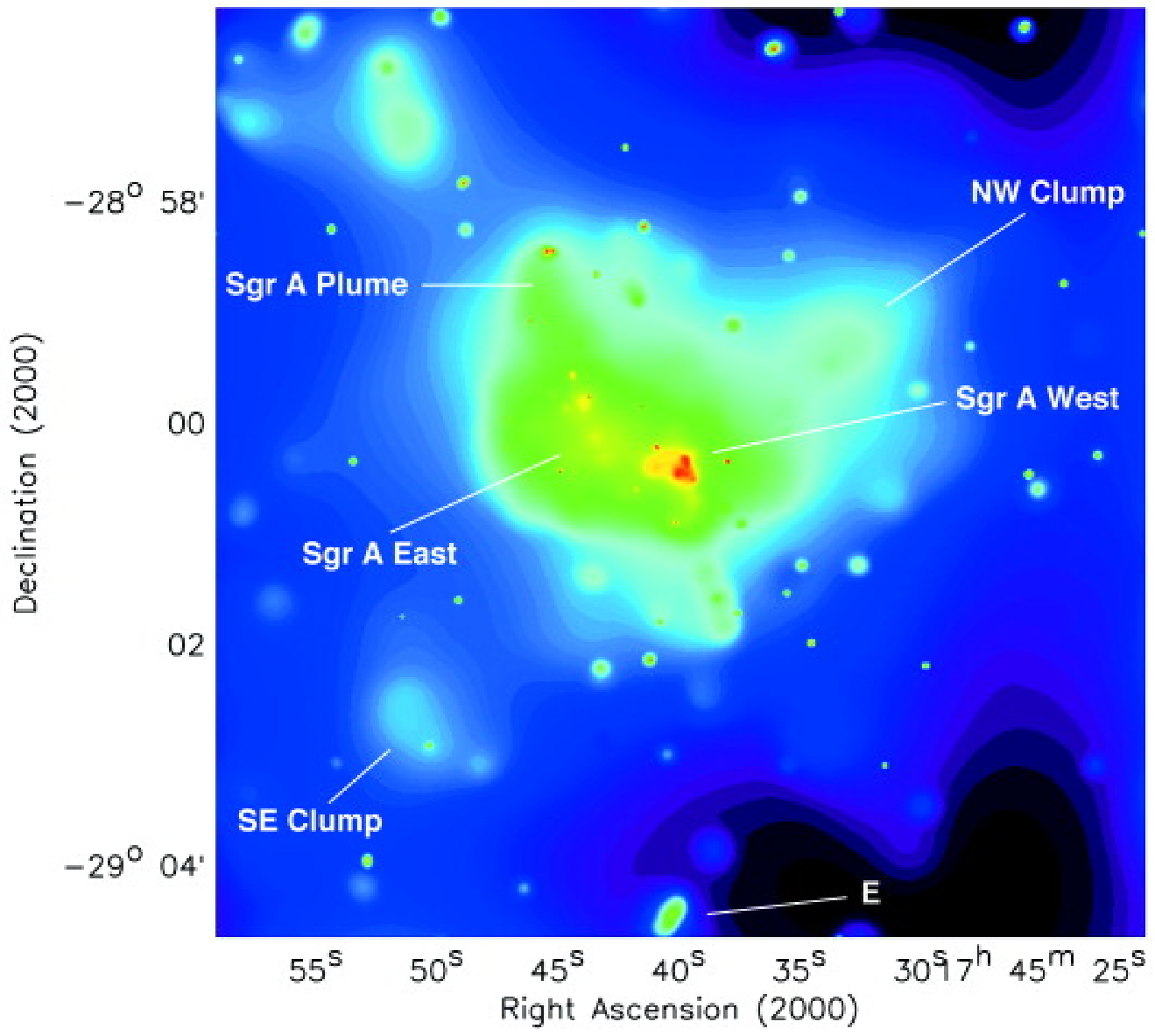}{2.5cm}{0}{44}{44}{10pt}{0pt}
\caption{
\footnotesize
Chandra raw image of the inner 1$'$ (2.4~pc) of the Galaxy (left)  
and the smoothed image of the inner 8$'$ (20~pc) (right) in the 2-10 keV band
\cite{bag03}.
}
\label{Chandraimage}
\end{figure}

\subsection{Flaring activity of Sgr~A$^*$}
One of the major results of the Chandra monitoring of the GC 
has been the discovery of X-ray flares from \sgra\,
one year after the detection of the quiescent emission \cite{bag01}. 
Follow up monitoring both with Chandra
and XMM-Newton have, since then, led to detection of several such events 
(Goldwurm et al. 2003; B{\' e}langer et al. 2005; Porquet et al. 2003, 2008; 
Trap et al. 2010, 2010b).
These flares typically occur ramdomly once a day, last 1 to 3 hours, and show 
a flux increase by factors up to 160 times the quiescent value,
with the most bright ones reaching luminosities of $\sim$ 3~10$^{35}$~erg~s$^{-1}$.
The spectra do not seem to vary during the event and usually display 
a harder spectral slope then the quiescent emission even if 
the brightest ones (and the most precisely measured) appear softer.
The flare duration and the observed short-time-scale 
variations, some of them as short as to 200~s, indicate that the X-ray emission 
is produced within 20~R$_S$. These flares allow us then to explore the inner 
region of the accretion flow and have been investigated deeply 
since their discovery.
They cannot be accounted for by the standard RIAF models, 
for which the X-ray emission is produced from the
whole accretion flow starting at the accretion radius, and 
clearly a variable, non-thermal component plays a major role during 
the flaring states. Whether this component in generated in a small hot
accretion disk \cite{liu04}, at the base of a compact jet \cite{mar01} or in the inner region
of a RIAF accretion flow \cite{yua04} is still matter of debate \cite{yua10}. 
Most of these models can account for the observed flare X-ray spectral shapes, 
but they predict different correlations between fluxes at different frequencies.

For these reasons, and after the discovery in 2003 
of near-infrared (NIR) flares from \sgra\ \cite{gen03,ghe04}, large efforts have been devoted 
in the recent years to carry out large multiwavelength (MWL) campaigns 
(from radio to gamma-rays) on \sgra\ 
(Eckart et al. 2004, 2006, Yusef-Zadeh et al. 2006, 2009, Hornstein et al. 2007, 
Marrone et al. 2008, Dodds-Eden et al. 2009). 
While NIR emission in \sgra\ flares is nearly certainly produced by synchrotron emission,
it is still not clear whether X-rays are originated by inverse Compton (IC) or synchrotron mechanism
as several recent results have shown that IC may imply extreme parameters for the emitting
region \cite{dod09,tra10}.
Moreover delays of radio and sub-mm flares with respect to NIR or X-ray events 
have been interpreted as due adiabatic expansion of the emitting plasma \cite{yus06b}. 
The large MWL campaigns are therefore now devoted also to test the expansion 
paradigm and set constraints on such process (Trap et al. 2010b).

Another essential, but also somehow controversial result on \sgra\ flares is 
the possible presence of quasi-periodic modulation on timescales of about 20 minutes,
which, if confirmed, would favor disk over jet models for \sgra.
Initially observed in the NIR data \cite{gen03} this feature was then reported also
in X-ray flares \cite{asc04,bel06b}.
If such a modulation is real and is associated to orbital motion at the last stable orbit 
of the accretion disk, then the observed timescale implies 
that the SMBH is rotating with a spin parameter of $\sim$~0.2 or higher. 
However several recent studies do not confirm 
the previously detected QPO signals in IR \cite{mey08,do09} and X-ray \cite{bel06b} 
light-curves when red-noise and all other effects are properly considered. 
In spite of other NIR hints of modulation both in flux and in linear polarization 
(Eckart et al. 2006b, Dodds-Eden et al. 2009) the matter is not settled and investigations 
are in progress.
For what concern the X-ray data, all recent observations of X-ray flares, 
including those in 2007 and 2009 with XMM, do not show quasi-periodic  
modulation in the light curves.

\subsection{The inner 20 pc: Sgr~A~East}

Moving from \sgra\ outwards, the morphology of the X-ray emission clearly 
shows an asimmetry towards positive galactic longitudes,  
with a bright diffuse oval source associated to the radio source Sgr~A~East
(Fig.~3 right). 
The detailed analysis of this emission with Chandra and XMM
has clarified the nature of Sgr~A East, which appears now to be
a single, compact, mixed-morphology SNR, where the non-thermal radio shell 
surrounds a centrally peaked thermal X-ray emission \cite{mae02}. 
The X-ray plasma has 2 components, one at 1~keV 
and the other at 4~keV \cite{sak04}.
High element abundances (Z $\sim$ 4) in the center of the source indicate 
that part of the emission is due to heated SN ejecta. 
The X-ray data show now that Sgr~A~East,
apart from its high plasma temperature and the dense medium where it expands, 
is not an exceptional SNR, nor the result of several SN explosions
or of the explosive disruption of a star by the SMBH as speculated previously. 
It rather appears to be the product of a typical SN II or a SN Ia occurred 
about 10$^4$~yr ago or less. The most recent investigations 
with Chandra \cite{par05} and Suzaku \cite{koy07b} have basically confirmed
all this, with a supporting evidence for a SN II origin and also the detection
of an additional power-law component. This non-thermal emission
could be due either to the cumulated contribution of the point source 
population detected with Chandra (see below) or to
a genuine non-thermal X-ray emission generated by the SNR.
Some authors have speculated that
the Sgr A East shell of swept up interstellar matter 
could have reached the SMBH feeding it and triggering
the Sgr~A$^*$ outburst of hard emission that is now illuminating Sgr~B2 \cite{mae02}.
However the role of Sgr A East in feeding the black hole with swept up material 
have been questioned in favour of a picture where a GC past X-ray outburst 
was rather generated by the interaction of the SNR shell with dense material
of the 50 km/s molecular cloud \cite{fry06}.

\subsection{The X-ray discrete source population}

The Chandra monitoring of the GC with a total exposure of 2.25 Ms led to the detection of 
over 9000 discrete sources in the 2$^\circ$ $\times$ 0.8$^\circ$ region around the nucleus
\cite{mun09}.
These sources appear spatially distributed as the stellar population, which is dominated
by old stars of the inner galactic bulge, but with density enhancements correspondent 
to the 3 young star clusters of the region, the Arches, the Quintuplet and the 
Galactic Center clusters. While in these star forming regions  
young high-mass X-ray binaries, pulsars, or Wolf-Rayet/O stars in colliding-wind binaries 
may contribute significantly to the population, the bulk of the weak
sources, with luminositites between 10$^{31}$-10$^{33}$~erg~s$^{-1}$, 
are probably cataclismic variables (CV) with a good fraction made of the 
hard magnetically-accreting white dwarfs.

The brightest sources, with luminosities larger than 10$^{34}$~erg~s$^{-1}$
are mostly low mass X-ray binaries,
like the persistent sources 1E~1740.7--2942 and 1E 1743.1--2843,
some are transients and some display type I bursts indicating that the compact 
object is a neutron star (e.g. GRS 1741.9-2853, see Trap et al. 2009). 
A total of 19 such XRBs have been detected in the GC region
since the beginning of X-ray astronomy
with 8 of them discovered by Chandra, XMM or Swift in the last 
10 years \cite{mun09}.
%
More than 20 diffuse non-thermal features, identified in the central 40 pc and
not associated with known objects, have been interpreted 
as PWN \cite{mun08}. These objects may provide important contribution to the gamma-ray emission
as we know that they are sites of particle acceleration.
The peculiar PWN G09+01 that is also a shell SNR and a prominent extended object in X-rays 
is indeed the second brightest source of the region at TeV energies \cite{aha06}.

\subsection{The hot component of the X-ray diffuse emission}

The diffuse X-ray emission of the galactic center region is complex and 
still under intense investigation but it certainly consists of at least three components
(Muno et al. 2004, Park et al. 2004, Koyama et al. 2009): 
a patchy soft emission well described by a low temperature ($\sim$1 keV) plasma model, 
a more uniform 6.7 keV line associated to a continuum emission described by a hot 
(kT $\sim 7$ keV) plasma, 
and a clumpy 6.4 keV iron line component well correlated to molecular material.
The soft component, traced by low-energy ($<$ 3~keV) K$_{\alpha}$ and K$_{\beta}$ 
lines of He-like and H-like ions of Si, S, Ar, Ca, and in particular 
by the strong He-like K line of sulphur at 2.46~keV, can be fully explained by 
the interaction of supernova remnants or stellar winds from young and 
massive stars with the interstellar matter of the region. 
The estimated SN rate and the concentration of this component towards the star 
forming regions 
close to the radio arc and the Arches culster are indeed consistent with 
these interpretations. The origin of the other components is more uncertain.

The hot component, characterised by the strong He-like (6.7~keV) and H-like (6.9~keV) 
ionized iron lines, is uniformely distributed, concentrated along the galactic plane 
with a peak towards the center and it is very similar to the GRXE.
The interpretation of a diffuse hot plasma emission with temperature of 6-7~keV 
is problematic because such a hot gas cannot be confined in the region and it would 
escape in $<$ 10$^5$~yr. Its regeneration would require a too large amount of energy
and an unknown source of power. 
Some authors have proposed that the gas be dominated by helium 
rather than hydrogen, in which case it would be bound to the region \cite{belm05}.
The motion of molecular clouds through the strongly magnetized medium of the GC
could then provide a heating mechanism, by dissipation of hydromagnetic waves
energy. 
Another possible explanation is that the GC hot component is simply produced 
by an unresolved population of weak discrete sources, presumably CV \cite{wan02}. 
Revnivtsev et al. (2007, 2009) have indeed shown that this is the case for the 
GRXE.
These authors have analyzed a Chandra galactic ridge ultra-deep field and found that 
as much as 88$\%$ of the diffuse 6.7~keV emission is explained by discrete sources,
mostly accreting white dwarf with 2-10~keV luminosity in the range 
10$^{31}$-10$^{32}$~erg~s$^{-1}$ or coronally active binary stars with lower luminosities.
A similar situation could hold for the GC, 
and this interpretation is supported by the similarity of the hot component spectrum 
with the cumulated spectra of the GC weak point sources observed with Chandra
\cite{mun04}. 
The point-source population detected with Chandra 
can account only for a fraction of the total diffuse emission of the GC region,
from 10-20$\%$ in the central area \cite{mun04} to 40$\%$ in a region centered at
negative longitudes \cite{rev07}. 
In order to explain the whole hot diffuse component the luminosity function 
of point sources would have to be extrapolated downward by few orders of magnitude. 
Moreover the spatial distribution of the diffuse component does not seem to be
fully compatible with the smooth one expected from point sources \cite{mun04}. 

Recent results obtained with Suzaku (Koyama et al. 2007, 2009), through a
detailed spectral study of the diffuse emission lines of ionized iron and nickel, 
show that the line centroids, widths and flux ratios 
favor a collisional excitation plasma in ionization equilibrium with temperature
of 6.5 keV.
Suzaku measurements also show that the line emission is much less prominent 
along negative longitudes unlike the distribution of weak discrete sources.
These facts seem to support again the hypothesis that the GC hot diffuse emission 
contains an important fraction of truly diffuse component but the issue is 
still higly debated.

\subsection{The 6.4 keV line diffuse emission}

The other distinct component of the GC X-ray diffuse emission 
is the 6.4~keV fluorescence line of neutral or weakly-ionized iron. 
This has a different morphology than the 6.7~keV line, being much less uniform, clumpy 
and clearly correlated to the molecular material of the Central Molecular Zone (CMZ).
The fluorescence line at 6.4 keV (K$_{\alpha}$ of Fe) is produced by the emission
of a photon that follows the extraction of an
electron from the inner shell (K) of neutral iron atoms as the result of
the electron transition from the second shell (L). 
Collisionally-ionized iron atoms in a hot plasma would rather produce lines
in the 6.5-6.9 keV range, associated to a plasma continuum spectrum. 
Thus, the origin of the 6.4 keV line is certainly non-thermal and must be associated 
either to irradiation by photons with energies higher than 7.1 keV \cite{sunchu98}
or interaction of energetic particles, most probably low energy electrons \cite{val00}. 
In the case of photoionization, a K edge absorption feature at energies higher 
than 7.1 keV is produced in the underline continuum. 
In both cases a continuum emission should be associated to the line: 
in the case photo-ionization it is due to the Thomson scattering 
of the incident radiation by the electrons of the cold material 
while in the case of interacting particles it is produced mainly
by non-thermal bremsstrahlung of the energetic electrons.
In case of irradiation and when the primary source is not contributing to the continuum
the line should display a large equivalent width ($\sim$~1~keV). 
The main feature in the GC 6.4~keV images is Sgr~B2, which
is the densest (core densities of 10$^5$-10$^6$ cm$^{-3}$) 
and most massive ($\sim$10$^6$ M$_\odot$) of the MC of the galaxy,
and is located at about 100~pc in projection from the nucleus.
Originally observed with ASCA \cite{koy96,mur00} 
it was later explored with Chandra \cite{mur01} and recently with Suzaku \cite{koy09}.
The emission was interpreted as fluorescent line due to scattering of 
hard X-ray emission coming from an external source, possibly a 
strong outburst ($\sim$ 10$^{39}$ erg s$^{-1}$) of hard X-rays 
from \sgra\ occurred some 300~yr ago
that have traveled the distance to Sgr~B2, illuminating the dense cloud and 
generating the Fe line and the reflected hard X-ray emission.
Hard ($>$ 10 keV) X-ray emission from Sgr B2 has been 
clearly observed up to 200 keV with INTEGRAL \cite{rev04} supporting 
the model of an X-ray reflection nebula (sect. 4).
These authors also demonstrated that the emission line intensity has been constant 
till about 2000 and therefore that the original outburst must have lasted at
least 10 years, excluding an X-ray binary as possible primary source.
The Fe K$_{\alpha}$ line has been detected in other molecular clouds 
of the CMZ, like Sgr~C \cite{mur01b,nak09} and G0.1-0.1 
(Yusef-Zadeh et al. 2002)
but not in all and a correlation of certain 6.4~keV hot spots 
with non-thermal radio filaments, that indicate presence of accelerated electrons,
has been found with Chandra \cite{yus07}. 
These elements seem to favour the alternative interpretation for the 6.4 keV line origin, 
namely the iron excitation by subrelativistic particles, 
either electrons (Valinia et al. 2000; Yusef-Zadeh et al. 2002, 2007),
or protons \cite{dog09}.

Although Suzaku with its high spectral resolution and low background level
has provided more precise spectral measurements of the GC Fe K$_{\alpha}$ line
\cite{koy09,nak09}, the most convincing evidences that 
support the photon-ionisation model from an external source, come now from the 
recent detections of variability of the line and associated continuum, 
a signature predicted and modelled in detail by Sunyaev $\&$ Churazov (1998).
First indication of local variability in the X-ray continuum (not in the line) of MC located
between \sgra\ and G0.1-0.1 were reported using Chandra data \cite{mun07}. Then, 
using archival data of several satellites (ASCA, XMM, Chandra) and Suzaku
measurements, it has been shown that Sgr B2 Fe K$_{\alpha}$ emission
is changing as it would be produced by a wave front passing trough the different components 
of the molecular complex \cite{koy08,inu09}.
But new extremely compelling evidences are given now by two crucial measurements.
The first is the significant detection of time evolution of the hard X-ray emission 
from Sgr B2 observed over 7 years by the same instrument on the INTEGRAL satellite 
(Terrier et al. 2009, 2010). The decrease in the 20-60 keV flux
of Sgr B2 is compatible with the reported decay of the 6.4 keV line intensity and 
best explained by a X-ray reflection nebula scenario where the fading of the 
reflection component 
is due to the propagation of the outburst decay through the molecular complex.
Even more stricking is the discovery of variations in the 6.4~keV line flux and morphology 
from the MC around 15$'$ from \sgra\ between 2002 and 2009 obtained  
by Ponti et al. (2010) using the large XMM database of the GC.
These authors report a significant decrease of G0.1-0.1 flux, an increase with
apparent superluminal propagation of the emission along a molecular feature called "bridge" 
and a constant behavior of other MC of the region. 
Such a variability and in particular the superluminal propagation 
exclude the particle model as particles cannot produce such rapid changes. 
On the other hand, for a X-ray nebula scenario, the constraints on the illuminating 
primary source indicate \sgra\ as the most probable candidate.
Using a new localisation of Sgr B2 along the line of sight and assuming a cloud 
distribution that is coherent with radio molecular line data 
these authors also show that the complex pattern of variations, including the decay in Sgr B2, 
can be due to one single outburst of \sgra\ that started 400 years back 
and ended about 100 years ago. 
The beginning of the burst is switching on the bridge, while its end has reached both 
Sgr B2 and G0.1-0.1. 
While the \sgra\ sigle outburst is not fully demonstrated by the data, 
these measurements clearly indicate that the galactic supermassif black hole 
has been bright, for few hundred years, with a luminosity of 
about 10$^{-5}$ times its Eddington luminosity till about 100 years back. 
Thus \sgra\ was remarkably more similar to typical low-luminosity AGN in the 
recent past than it appears today.
\begin{figure}[!ht]
\plotfiddle{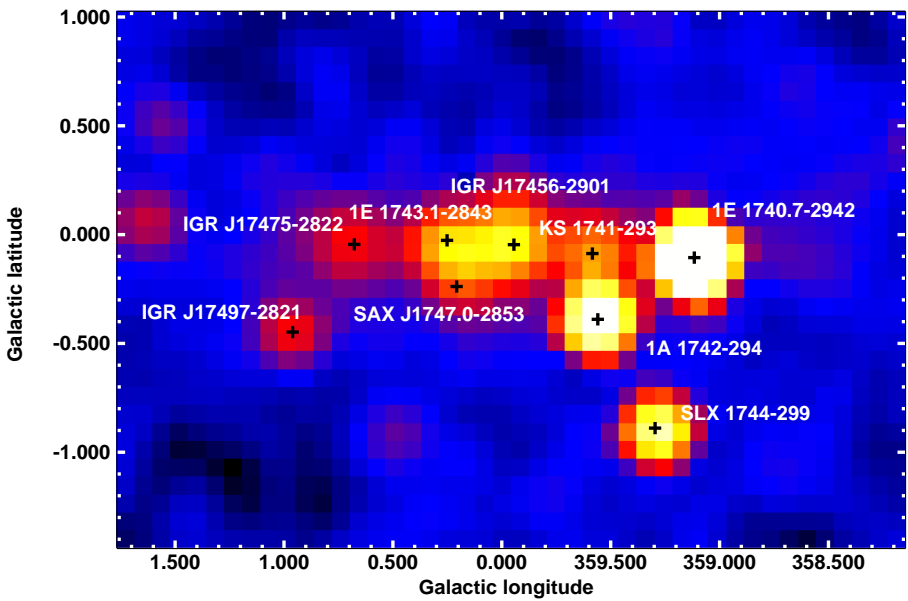}{7.8cm}{0}{73}{73}{-210pt}{-180pt}
\caption{
\footnotesize
INTEGRAL image of the galactic center in the 20-40 keV band
from all IBIS/ISGRI data collected between 2003 and 2009 
for 20~Ms of exposure time \cite{ter10}.
}
\label{ibisimage}
\end{figure}

\section{The Galactic Center in gamma-rays}

The GC emission at energies $>$~20~keV has been deeply explored in the last 10 years  
with INTEGRAL, HESS and more recently with FERMI.
INTEGRAL has monitored the GC for more than 20~Ms between 2003 and 2009, 
obtaining with the IBIS/ISGRI telescope the most precise GC images ever collected 
in the 20-800~keV band \cite{bel06,ter10} (Fig. 5).
In this band the emission is dominated by bright and variable XRB
with the hardest one beeing the BH microquasar 1E~1740.7-2942, 
but INTEGRAL also detected 
a faint and persistent emission coming from the very center of the Galaxy, 
compatible with a source located within 1$'$ of the \sgra\ position.
Due to the IBIS angular resolution ($\sim$ 13$'$ FWHM) this source
(IGR~J17456--2901) cannot be identified with the SMBH or 
other objects of the dense central region.
The lack of variability and of a bright discrete X-ray counterpart
suggests that it is rather a diffuse emission concetrated 
in the inner central 10-20 pc \cite{bel06} or the sum of 
the contribution of unresolved point-like sources \cite{kri07}.
The spectrum at $>$~20~keV cannot be explained by the extrapolation of the thermal 
plasma with kT of 6.5 keV used to model the bulk of the X-ray diffuse emission 
and a non-thermal component extending up to 150 keV with steep spectral slope
is clearly present and its origin is still unexplained.
While the low-energy emission can be understood as the sum of unresolved weak discrete sources 
even the hard Intermediate Polars CV are not hard enough to explain the observed
signal at $>$~100~keV. However the high-energy part of the spectrum ($>$~50~keV) could be  
be produced by closeby MC 
and this would explain the slight displacement observed in the centroid of IGR J17456--2901 
at high energies.
The presence of hard ($>$~10 keV) non-thermal emission from the GC was later confirmed
with Suzaku \cite{yuas08} which also showed that it could be more extended
than the central 20~pc.
Simultaneous XMM-INTEGRAL observations performed during the 2007 
brigh \sgra\ X-ray flare, set upper limits 
on the variable component of the gamma-ray ($>$ 20~keV) emission of IGR
constraining somehow the \sgra\ flare mechanisms \cite{tra10}
and confirming that these flares do not contribute significantly 
to the central soft gamma-ray emission.
In addition, INTEGRAL detected hard X-rays
from Sgr B2 (IGR~J17475-2822) \cite{rev04}, and especially, its significant
decay between 2003 and 2009 (Terrier et al. 2009, 2010), fully compatible, 
as discussed in sect.~3.6, with the decay observed in the 
6.4~keV line and with the hypothesis that Sgr B2 is a reflection nebula,
scattering radiation generated by \sgra\ in the past.

Concerning the 511~keV line emission, the spectrometer on INTEGRAL has now mapped
the sky with the highest accuracy available 
and has shown that this emission is extended over the galactic bulge ($\sim $8$^{\circ}$ FWHM), 
is composed by both a bulge and a weaker disk components, 
and that the positron-electron annihilation take place partly in a warm neutral gas and 
partly in a warm ionized medium.
The origin of the bulk of the positrons however is not yet understood 
(see Skinner et al. 2008 for a recent review). 
While the favorite mechanisms are not generally located in the very center of the Galaxy, 
a possible role of the SMBH in generating positrons has been investigated recently
\cite{che06,tot06}.

In the VHE gamma-ray band the new generation of Atmospheric Cherenkov Detectors (ACD)
have provided spectacular results on the GC. 
HESS, the most sensitive and precise of them, reported 
(after the first detections with Whipple and Cangaroo) the presence of
a bright TeV point-like source centered within 1$'$ from \sgra\ (Aharonian et al. 2004, 2009).
The source is constant and displays a power-law spectrum extending from 300 GeV 
up to $>$ 10 TeV with a break around 15 TeV.
It cannot be explained by heavy dark matter particle annihilation, 
because the spectrum is too hard and would imply too massive particles.
It is rather attribued to interaction of accelerated leptons or hadrons,
but the mechanism, site of acceleration and nature of primary particles are not yet identified.
The expanding shell of the Sgr~A~East SNR, once considered as the 
best candidate for the TeV central source \cite{cro05},  
it is now formally excluded by the most recent HESS determinations of the location 
and error box of the source, which appear well centered on \sgra\ and too far from 
Sgr~A~East centroid \cite{ace10}.
A serious candidate for HESS~J1745--290 is instead 
the energetic cometary-like pulsar wind nebula (G359.95--0.04) (sect. 3.1)
that could contribute to the TeV emission by IC scattering of the strong ambient 
radiation by the electrons accelerated in the nebula.
Otherwise the SMBH is still a potential candidate \cite{ahaner05} even if 
the GC HESS observations during a \sgra\ flare observed with Chandra
have shown that the central TeV source did not vary 
during the eruption \cite{aha08} and does not display variability 
on any time scale.
Even if HESS~J1745-290 is not directly connected to \sgra\ flares, 
particles accelerated in the inner regions close to the BH could 
propagate and interact with the surrounding matter
close enough to produce a point-like source for the present ACD telescopes.
The second brightest source of the region at TeV energies is the composite PWN and shell SNR
G09+01 but the HESS collaboration has also reported the discovery of 
TeV diffuse emission closely correlated to the molecular clouds of the CMZ \cite{aha06}. 
The spectrum and distribution of this emission seem consistent with 
the idea that the central source (which has similar spectrum) accelerated in a recent past 
(few kilo-years) the hadrons that diffused in the region interacting with the molecular gas. 
Other studies \cite{wom08} seem to show that, in order to explain 
the smooth distribution of the emission, the origin of the hadrons cannot be the central 
source nor a distribution of point-like objects (e.g. pulsar wind nebulae).
Those authors concluded that only a relativistic proton distribution accelerated 
throughout the intercloud medium can account for the TeV emission profile measured 
with HESS. 

In the gamma-ray domain of medium-high energies the FERMI observatory, launched in June 2008 
and working in the range 50 MeV - 100 GeV, has confirmed the EGRET results by detecting 
a source close to \sgra, listed in the FERMI/LAT catalogue \cite{abd10}
as 1FGL~J1745.6--2900c at a position 
compatible with several possible candidates including the SMBH, the PWN 
G359.95--0.04 and Sgr~A East. No variability has been reported untill now for this source.
It is interesting to note that another significant excess (1FGL~J1747.6--2820c)
compatible in position with Sgr B2 is also listed in the FERMI catalogue while EGRET did not detect 
any excess emission in Sgr B2 \cite{may98}.
Obviously a specific, thourough analysis of FERMI data is needed in this complex and confused region 
of sky to measure the effective excess emission with respect to the one that is expected 
from cosmic ray (CR) interaction with the dense interstellar matter.
Indeed the gamma-ray measurements can indicate whether the CR 
production and density in the GC is enhanced with respect to local or galactic disk values.
Several measurements seem in fact to indicate that the ionization rate in the CMZ, 
with an estimated value of 3~10$^{-16}$ s$^{-1}$ for Sgr~B2, is 10 times higher than 
in the galactic disk. 
However the additional CR component (factor 10) required by the GC TeV diffuse emission 
for the hadronic model has a hard spectrum, its gamma-ray contribution become negligible 
at $< 100$ GeV and cannot explain the additional ionization rate \cite{cro07} for which 
a specific steep low-energy component seems needed.
To fit all these elements toghether, including a possible lack of detection of Sgr B2 
in the GeV range and the lack of detection of non-thermal radio emission from the secondary leptons
produced by the hadronic interactions \cite{cro07}, more results are needed in particular 
in the FERMI energy band.

In conclusion: gamma-ray emission from the GC has now been clearly detected, 
however its origin and nature are not fully understood and several new questions arise.
The data that are being collected will certainly provide new results in the near future
and will possibly clarify some of the listed issues.

\section{Open questions and perspectives}

As discussed above, the topics related to the high-energy GC emission that are today 
still highly debated concern: the mechanisms of both quiescent and flaring emission of \sgra;
the characteristics of its activity in the recent past; 
the origin of the diffuse 6.7~keV and associated continuum emission 
(sum of discrete sources or genuine diffuse hot component); 
the nature of the 6.4~keV diffuse emission and the associated non-thermal continuum; 
the nature of the Integral, Hess and Fermi sources detected in the very center of the Galaxy;
the mechanism of production of the TeV and GeV diffuse emission and the 
origin and role of accelerated particles in the emission and ionization state of the region.

Future high-energy oservations coupled to MWL campaigns will probably 
elucidate the mechanisms of \sgra\ flares in the coming years and address 
some of the main issues on \sgra\ physics. 
These programs will possibly settle the issue concerning the presence 
of periodicities in the \sgra\ emission during flares and 
will provide more thourough measurements of the broad band spectra
of \sgra\ during these events up to 10 keV.
Chandra, XMM and Suzaku will continue to monitor the properties 
and the variability of the diffuse X-ray emission, testing the reflection nebula 
paradigm and exploring the nature of the hot diffuse component. 
FERMI will provide soon important new results on the central 
GeV source and Sgr B2 while the next generation of ACD 
(e.g. HESS~2 and the Cherenkov Telescope Array) 
will map in the future the GC at TeV energies with increased precision.
To finally solve the puzzle of the hard X-ray emission detected by INTEGRAL (and Suzaku)
at the very center of the Galaxy and to fully understand the related items 
it will be however necessary to wait for focusing instruments of hard X-rays.
With the stop of the Simbol-X mission \cite{fer08,gol08}, 
perspectives are less optimistic in this domain now.  
Nevertheless the data of Astro-H, NuSTAR and Spectrum-Roentgen-Gamma, 
which are expected to fly in few years, will certainly 
provide important contributions to address some of the topics concerning 
the origin of the hot gas, the properties of the GC hard 
source populations, the role of reflection in producing the 6.4 keV line and 
the hard X-rays in the GC region.


\end{document}